# Multilayer Graphene Synthesized by CVD Using Liquid Hexane as the Carbon Precursor

C. Bautista, D. Mendoza

Departamento de Materia Condensada y Criogenia, Instituto de Investigaciones en Materiales, Universidad Nacional Autónoma de México, Distrito Federal, México.
Email: claudiabautistaf@gmail.com, doroteo@servidor.unam.mx.

**ABSTRACT**

*We produce multilayer graphene by the Chemical Vapor Deposition (CVD) method at atmospheric pressure and 1000 °C, using flexible copper substrates as catalyst and liquid hexane as the source of carbon. We designed an optical device to measure the transmittance of the carbon films; with this information we calculate that the approximate number of layers is 11.*

## 1. Introduction

Since 2004, graphene has been proved to be a very interesting new material [1] because of its amazing optical, electrical and phononics, particular properties [2]. First of all, graphene is only one carbon thin layer with 0.34 nm thickness in a hexagonal arrangement, the so called honeycomb lattice. Electrons in graphene can travel trough the honeycomb lattice with a zero effective mass and without scattering, described by a 2D analog of the Dirac equation [3,4] with a Fermi velocity $v_F=1\times10^6$ m/s [5]. A half integer quantum Hall effect for both electron an hole carriers is observed in graphene [3, 6], and the Klein paradox was also experimentally observed [7]. One of the most important features of graphene, which is relevant to our research, is that graphene´s absorbance is a universal constant given by: $\pi\alpha=\pi/137=2.3\%$, where $\alpha$ is the fine structure constant [8-10]. We use this quantity to estimate the number of layers of graphene in our graphitic films. Due to its unique properties, graphene synthesis has acquired a great interest. Chemical Vapor Deposition (CVD) is a useful method to produce large area graphene films on metal surfaces [11]. With the CVD method, metals such as Ni [12], Co [13], Ru [14], Ir [15] and Cu [11, 12, 16-17] are used; and $CH_4$ or $C_2H_2$ as a source of carbon are employed [18,19]. Cu is the most used metal to produce large area graphene films [18-22]. Because of the versatility and cost, we prepared multilayer graphene on top of a Cu foil using hexane ($C_6H_{14}$) as a liquid precursor of carbon.

## 2. Experimental Details

We used Cu foils of 1.5x1.5 cm$^2$ and 100 μm of thickness to grow multilayer graphene, using liquid hexane as a source of carbon. Our CVD system, with a container for liquids, is very versatile; we can change without difficulty the liquid hexane by other liquid precursor. Therefore, growth of graphene using liquid precursor have some advantages over gas precursor; one of these is, gas precursor are more expensive than a liquid one, but one of the most important issue is that using liquid phase it would be possible to dope graphene [23]. The Cu foils were immersed in a ferric nitrate

(Fe(NO$_3$)$_2$-9H$_2$O) solution for a chemical polish during 2.5 hours to have a more uniform Cu surface; because carbon film take the morphology of the substrate in the CVD method, and we are interested in well structured few layer graphene.

Using a horizontal furnace, Cu foils were introduced at the center of a 70 cm long fused quartz tube with 2.5 cm internal diameter. We found that the better conditions to have the thinner films are the following: In a hydrogen atmosphere (473 sccm) the system go from room temperature to 1000 °C, at this time, we let to flow 4 sccm of H$_2$ into a bubbler containing liquid hexane during 15 minutes into the quartz tube, after this time the hexane flow is cut off and the furnace is turned off. The hydrogen flow is maintained until the system was cooled to room temperature. The CVD system is presented in Figure 1. After CVD process, Cu foils were etched using a ferric nitrate solution during the night, 100 mg of ferric nitrate in 20 ml of water. After this process, the carbon film is transferred to a deionized water bath to clean it, and finally captured with the desired substrate. Corning glass or fused quartz substrates for transmittance measurements were used, and copper grids for electron diffraction characterization.

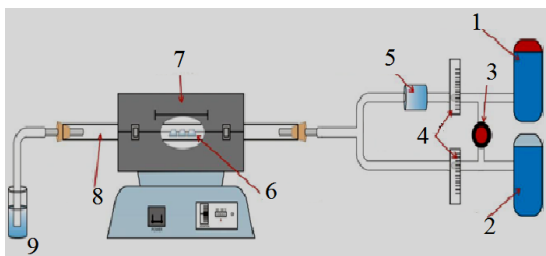

**Figure 1. CVD system: 1) Hydrogen gas, 2) Argon gas, 3) gas valve, 4) Flow controllers, 5) Hexane container (bubbler), 6) samples, 7) CVD furnace, 8) quartz tube, 9) Water.**

We know that graphene´s absorbance is of 2.3% and it is practically independent of wavelength in the visible-infrared region. Furthermore, the absorbance of multilayer graphene increases with the number of layers, each graphene layer increases absorbance by 2.3% [10]. We use this approach to determine the number of layers of our multilayer graphene produced by the CVD method using liquid hexane as a source of carbon. To measure the transmittance of our films, we propose the experimental arrangement shown in Figure 2 a).

We adapted a photodiode (Infrared Industries, serial number 9002) in one of the ocular lenses of an optical microscope (Iroscope MO-64), to measure the transmittance of our multilayer graphene with white light, and color filters of 620 nm and 905 nm. The used objective was 60X, the transmitted light by the substrate and the film on the substrate were measured using the photodiode.

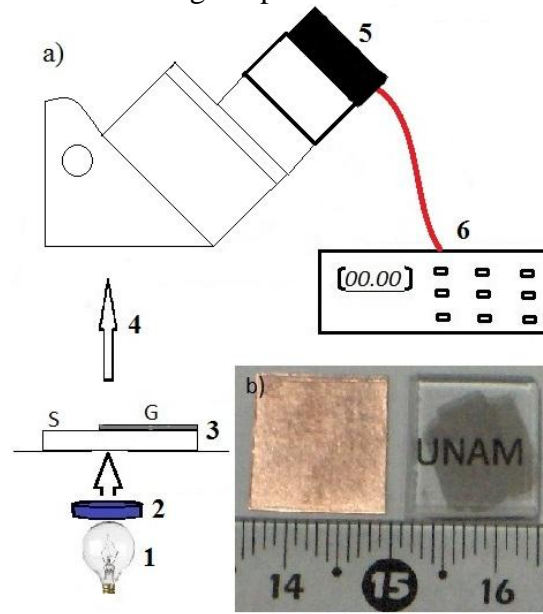

**Figure 2. a) Experimental arrangement for transmittance measurement: 1: light source, 2: filter, 3: sample where S means Substrate and G Graphene, 4: transmitted light collected by the objective of the microscope, 5: photodiode in place of the ocular, 6: electrometer. In b), samples obtained by CVD on Cu and fused quartz (right) for optical characterization.**

In this case, the current generated in the photodiode and measured by the electrometer is proportional to the light intensity.

## 3. Results and Discussion

The electron diffraction pattern of our carbon films obtained by CVD method is shown in Figure 3 a). We observed that when the sample is tilted respect to the electron beam, some diffraction spots change notably their intensities, such as those marked with arrows, indicating that we are dealing with multilayer graphene films [24].

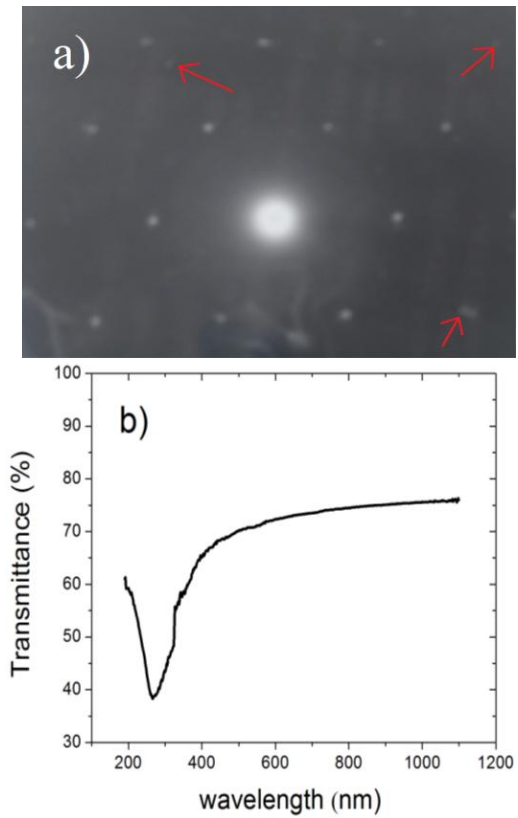

Figure 3. a) Electron diffraction pattern of our multilayer graphene using JEOL, JEM-1200EX at 120 KV. b) UV-Vis transmittance spectrum.

The UV-Vis transmittance spectrum of our films is shown in Figure 3 b) and it is similar to those reported for few layer graphene [19]. This spectrum shows an absorption peak around 268 nm, arising from resonant excitons [25], and in the visible-infrared range the transmittances is almost constant for a single layer graphene, but when the number of layers increases, the transmittance varies significantly in visible region, as in [19]. The measured transmittance of our samples obtained using the arrangement shown in Figure 2 is presented in Figure 4, we can see two important things, optical transmittance varies from white light to monochromatic light, and also between two specific wavelengths of incident light (620 nm and 905 nm).

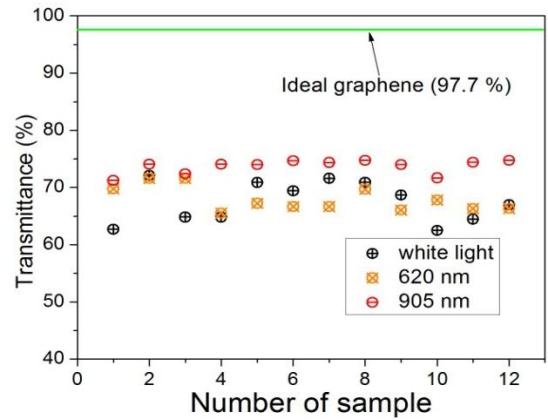

Figure 4. Transmittance obtained using white light and monochromatic light of 620 and 905 nm. Number of sample means different zones on the same macroscopic film, limited by the amplification generated by the 60x objective (300 μm of diameter approximately).

For the first one we can say that white light contains all wavelengths, and for the second one, the transmittance varies significantly when the number of layers increases, so we need to work only with monochromatic light near to the infrared region to have reliable results when we have multilayer graphene. If we generalize the use of a decrease in transmittance of 2.3 % per layer, the results for 905 nm in Figure 4 give an average of 73.70% of the transmittance, which is equivalent to 11 layers approximately.

On the other hand, we performed a homogeneity analysis of the films measured by their optical transmittance (see Figure 5a)) using the amount of red color in a digital image using the RGB system, such as the using in optical reflectance method [26]

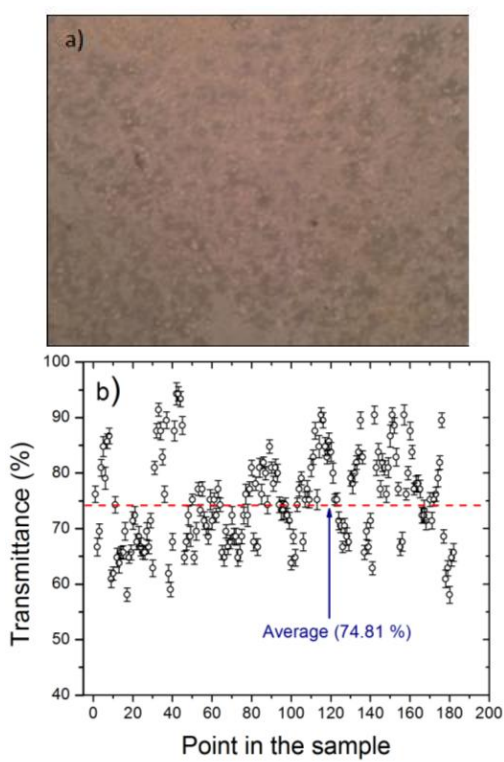

**Figure 5. Homogeneity analysis, in a) we show a digital image obtained by transmission of multilayer graphene films on top of a fused quartz substrate In b) the value of the transmittance obtained from the digital image is shown.**

The analysis was made point by point in the image of the substrate and the sample on top of the substrate using Corel Draw, the image was divided in a grid of 8x10 and in each point we took the amount of red color. Then, using the numbers generated by this software, the transmittance of the films can be calculated. With this analysis we have that our samples are homogeneous in 74.8% ±8.3% in transmittance and, generalizing the criteria of 2.3% of absorbance per layer, we can say that our samples consist mainly of 11 layers but with some zones with 7 up to 14 layers of graphene, approximately.

## 4. Conclusions

We obtained multilayer graphene using CVD method with liquid hexane as a carbon precursor at atmospheric pressure. We implemented a method to measure optical transmittance of our multilayer graphene and, with this information the number of layers in each sample is estimated, around 11 in our samples. We also generalized the analysis of digital images to calculate the transmittance of the films, point by point.

## 5. Acknowledgements

The authors wish to thank Miguel Ángel Canseco Martínez and Carlos Flores Morales by support in UV-Vis and TEM measurements respectively.